\documentclass[seceq,supplement]{ptptex}

\def\del{\delta}

\def\be{\begin{equation}}
\def\ee{\end{equation}}
\def\bea{\begin{eqnarray}}
\def\eea{\end{eqnarray}}
\def\la{\label}
\def\bsea{\begin{subeqnarray}}
\def\esea{\end{subeqnarray}}
\def\u{\underline}

\title{ A FDR-consistent field theory for the stochastic dynamic density functional model    
}

\author{
Bongsoo \textsc{Kim}$^{1,}$\footnote{Permanent Address: Physics Department, 
Changwon National University, Changwon 641-773, Korea, E-mail: bskim@changwon.ac.kr}
and Kyozi \textsc{Kawasaki}$^{2,}$\footnote{E-mail: tomo402000@yahoo.co.jp}
}

\inst{
$^1$Department of Theoretical Studies, Institute for Molecular Science, \\Okazaki 444-8585, Japan\\
$^2$ Electronics Research Laboratory, Fukuoka Institute of Technology, \\
Fukuoka 811-0295, Japan
}

\abst{
We present a brief review of our recent efforts to develop a FDR-preserving field
theory for the  stochastic dynamic density functional model, emphasizing the essential 
structure of the theory.
}

\begin{document}

\maketitle

In this proceeding paper we review our recent efforts of constructing a field theory consistent with 
the fluctuation-dissipation relation (FDR) for the stochastic dynamic density functional model which can describe
dynamics of colloidal systems (without hydrodynamic interaction) as well as classical atomic liquids.
Such a field theory can provide a systematic theoretical basis for the standard MCT \cite{reviews} which involves
totally uncontrolled approximations such as factorization of the $4$-body density correlations, and 
hence can offer a viable possibility of systematically improving the standard MCT.
Here we hope to elucidate the essential structure of our theory. For further details we refer the reader 
to our recent papers \cite{bkkk0708}.  

Our starting point is  the following stochastic time evolution equation for the density field $\rho({\bf r},t)$  
\be
\partial_t \rho({\bf r},t)=\nabla \cdot \Big( \rho({\bf r},t) \nabla \frac{\delta F[\rho]}
{\delta \rho({\bf r},t)} \Big) +\eta({\bf r},t)
\la{eqn:2.1}
\ee
where the Gaussian thermal noise $\eta({\bf r},t)$ has zero mean and the variance 
\be
<\eta({\bf r},t) \eta ({\bf r}',t')> = -2T \nabla \cdot \Big[  \rho({\bf r},t)\nabla \delta({\bf r}-{\bf r}') \Big]\delta(t-t')
 \la{eqn:2.2}
\ee
where $T$ is the temperature of the system, and the Boltzmann constant $k_B$ is set to unity.
 In (\ref{eqn:2.1}), $F[\rho]$ is the effective free energy density functional
which takes the following form:
\be
F[\rho]= T \int d{\bf r} \,
 \rho ({\bf r}) \Big(\ln \frac{\rho ({\bf r})}{\rho_0}-1 \Big) 
+ \frac{1}{2} \int  d {\bf r} \int  d {\bf r}' \, \delta \rho ({\bf r})\, U({\bf r}-{\bf r}')\,  \delta \rho ({\bf r}')
\la{eqn:2.3}
\ee
where $\delta \rho ({\bf r},t) \equiv \rho({\bf r},t)-\rho_0$
is the density fluctuation around the equilibrium density $\rho_0$.
In (\ref{eqn:2.3}) the first term is the entropic contribution $F_{id}[\rho]$, and the second one 
the interaction contribution $F_{int}[\rho]$. 

The above stochastic model (\ref{eqn:2.1})-(\ref{eqn:2.3}) was proposed by one of the present authors \cite{kk94} as 
a mesoscopic kinetic equation which can describe the long-time dynamics of both colloidal particles
and classical liquids. The equation for the coarse-grained density $\rho({\bf r},t)$ 
was obtained (with $U({\bf r})$ replaced by $-Tc({\bf r})$, $c({\bf r})$ being the direct
correlation function of the system) via adiabatic elimination of the fast-decaying momentum field 
in the fluctuating hydrodynamic equations \cite{dasmazenko} of dense liquids. 
For this case, (\ref{eqn:2.3}) takes the Ramakrishnan-Yussouff (RY) form \cite{ry}. 
The same form of the exact Langevin equation was derived \cite{dean} for the microscopic density 
$\hat \rho({\bf r},t)$ of Brownian particles with bare interaction potential $U({\bf r})$.
For further discussions regarding the nature (and controversy) of  the dynamic equation 
(\ref{eqn:2.1})-(\ref{eqn:2.3}), we refer to Ref. \cite{kk06}.

The  Fokker-Planck (FP) equation corresponding to (\ref{eqn:2.1}) and (\ref{eqn:2.2}) can be written as
\be
\frac{\partial P( [\rho],t)}{\partial t}=-\int d{\bf r} \frac{\delta}{\delta \rho({\bf r})}
\nabla \cdot \rho ({\bf r}) \Big[ T \frac{\delta}{\delta \rho({\bf r})}
+\frac{\delta F[\rho]}{\delta \rho({\bf r})}  \Big] P( [\rho],t)
\la{eqn:2.4}
\ee
where $P([\rho],t) $ is the probability distribution of the density variable $\rho({\bf r})$.
The equilibrium distribution $P_e [\rho] \propto \exp \big(-F [\rho] /T \big)$ 
is a (possibly the only) stationary solution of the FP equation (\ref{eqn:2.4}), which is  
rendered  possible due to the prescription of the multiplicative noise correlation (\ref{eqn:2.2}). 

Note that the noninteracting part of the free energy gives rise to the linear diffusion for the density 
fluctuation
\be
\nabla \cdot \Big( \rho({\bf r},t) \nabla \frac{\delta F_{id}[\rho]}
{\delta \rho({\bf r},t)} \Big)=T \nabla^2 \rho
\la{eqn:2.5}
\ee
This is an interesting fact since the two nonlinear contributions, $F_{id}[\rho]$ and the extra factor 
of $\rho({\bf r},t)$ in (\ref{eqn:2.5}) generate the linear diffusion which is physically expected for 
the noninteracting particles. In order to generate physically correct linear diffusion for the system of 
noninteracting particles, these two elements of nonlinearity are intricately connected.
Actually, as discussed below, all the complications in the analysis of the original Langevin equation
(\ref{eqn:2.1})-(\ref{eqn:2.3}) come from these two ingredients:
\begin{itemize}
\item{Nonpolynomial nonlinear density denpendece of $F_{id}[\rho]$}
\item{Multiplicative noise structure, i.e.,  the extra factor of $\rho({\bf r},t)$}
\end{itemize}
It has been explicitly shown by Miyazaki and Reichman \cite{mr} that due to the second element
the response function takes an unusual form (see (\ref{eqn:2.9})), and that 
the FDR does not hold in a Martin-Siggia-Rose (MSR)-type \cite{msr} renormalized perturbation theory (RPT). 
Andreanov, Biroli, and Lefevre (ABL) \cite{abl} elucidated the origin of the incompatibility of the FDR
 with the usual RPT by focusing on a physical symmetry of the action integral, namely, 
the time-reversal (TR) symmetry. 
ABL pointed out that the origin of this inconsistency lies in the first ingredient, 
the nonpolynomial nonlinearity of $F_{id}[\rho]$ entering into the TR transformation of the fields 
leaving the action invariant. ABL then proposed the introduction of the conjugate pair of auxiliary fields 
to linearize the TR transformation, which guarantees preservation of the FDR order by order.
The present development outlined below is a modified version of ABL's auxiliary field method.

The action integral ${\cal S}[\rho, \hat \rho]$ 
\cite{msr} from which one can systematically obtain the correlation and response functions of
the density variable, is derived as 
\be
{\cal S} [[ \rho, \hat \rho ]] \equiv \int \, d {\bf r} \int dt \,
 \Big\{ i\hat \rho \Big[ \partial_t \rho -\nabla \cdot \Big( \rho \nabla 
\frac{\delta F[\rho]}{\delta \rho} \Big)  \Big] -T \rho ( \nabla \hat \rho)^2 \Big\}
 \la{eqn:2.6}
\ee
where the auxiliary field $\hat\rho$ is a real field,  and the last term involving the quadratic 
$\hat \rho $ comes from the average over the multiplicative thermal noise $\eta$. 
Here and elsewhere the functional dependence on $\rho$ as a function of ${\bf r}$ is denoted as $[ \rho ]$, 
but the functional dependence on $\rho$ as a function of ${\bf r}$ {\it and} $t$ is denoted as $[[ \rho ]]$.
 In deriving (\ref{eqn:2.6}), employing  the It\^{o} calculus  
makes the Jacobian of the transformation constant.  
 (The dynamic action of this form with the RY free energy functional was first written down in \cite{kk97}. 
Here we use the original form  (\ref{eqn:2.3}) for $F[\rho]$. ) 

The action (\ref{eqn:2.6}) becomes invariant when $\rho $ and $\hat\rho $ transform under TR as
\be
\quad \rho ({\bf r}, -t) =  \rho ({\bf r}, t),  \quad \quad
\hat \rho ({\bf r}, -t)  =  -\hat \rho ({\bf r}, t)
+ \frac{i}{T} \frac{\delta F[\rho]}{\delta \rho ({\bf r},t)}
\la{eqn:2.7}
\ee
The same action also becomes TR-invariant under another field transformation
\bea
\rho({\bf r},-t)&=&\rho({\bf r},t), \quad \quad 
\hat\rho({\bf r},-t)= \hat\rho({\bf r},t)+i h(\rho({\bf r},t)) \nonumber \\
\nabla \cdot ( \rho({\bf r},t) \nabla h(\rho({\bf r},t)))
 &\equiv&  \frac{1}{T}\partial_t \rho({\bf r},t)
\la{eqn:2.8}
\eea
Note that both  (\ref{eqn:2.7}) and (\ref{eqn:2.8}) are intrinsically nonlinear transformations: 
the first one is nonlinear owing to the noninteracting contribution $F_{id}[\rho]$, whereas the second one 
due to the multiplicative nature  of the original Langevin equation. 
As elucidated by ABL \cite{abl},  the nonlinear nature of these transformations is the underlying reason 
why the FDR, obeyed by the action, is not preserved order by order in the 
RPT developed for the action (\ref{eqn:2.6}). 

The FDR  \cite{deker} is a fundamental relationship between the equilibrium fluctuations and the linear response 
to external perturbation. The response function $R({\bf r},t; {\bf r}' t')$, describing how much the 
average density $\big<\rho({\bf r},t)\big>$ changes under application of an 
external  infinitesimal field  $h_e({\bf r}',t')$ (added to $F[\rho]$), is given by 
the following form \cite{mr}
\be
R({\bf r},t; {\bf r}' t') =i\Big<  \rho( {\bf r},t) \nabla' 
\cdot \Big( \rho( {\bf r}',t') \nabla' {\hat \rho}( {\bf r}',t') \Big) \Big>
\la{eqn:2.9}
\ee
The unconventional form due to the extra factor $\rho( {\bf r}',t')$ in (\ref{eqn:2.9}) reflects 
the multiplicative nature of the original Langevin equation (\ref{eqn:2.1}) and (\ref{eqn:2.2}).
The FDR  then easily follows from the action (\ref{eqn:2.6}) and the TR transformation 
(\ref{eqn:2.7}) using an identity 
$\big< \rho ({\bf r},t) \big( \delta {\cal S}[[ \rho, \hat \rho]]
/\delta {\hat \rho}({\bf r}',t')\big) \big>=0$:
\be
-\frac{1}{T} \partial_{t} C({\bf r}-{\bf r}', t-t')
=-R({\bf r}-{\bf r}', t'-t) + R({\bf r}-{\bf r}', t-t')
\la{eqn:2.10}
\ee
where $C({\bf r}-{\bf r}', t-t')\equiv \big< \delta \rho ({\bf r},t) \delta \rho ({\bf r}',t')  \big>$ 
is the density correlation function. 
The FDR (\ref{eqn:2.10}) is more directly obtained from the second transformation (\ref{eqn:2.8}).
Therefore the original Langevin equation is consistent with the FDR.

Though the original equation is compatible with the FDR, 
the RPT developed from the action (\ref{eqn:2.6}) was shown to be incompatible with the FDR \cite{mr}. 
It was  subsequently shown \cite{abl} that the intrinsic nonlinear nature of the TR transformations 
(\ref{eqn:2.7}) and (\ref{eqn:2.8}) gives rise to this inconsistency. 
In order to resolve this inconsistency, it is thus crucial to develop a method which 
can properly linearize the nonlinear TR transformations. 

From now on, we focus on the first TR transformation (\ref{eqn:2.7}). 
With the form of the free energy given in (\ref{eqn:2.3}), 
the transformation (\ref{eqn:2.7}) is explicitly written as
\be
\rho ({\bf r}, -t) =  \rho ({\bf r}, t), \quad \quad
\hat \rho ({\bf r}, -t) =  -\hat \rho ({\bf r}, t) 
+i{\hat K} \ast \delta \rho({\bf r},t)+i\, {\cal F}(\delta \rho({\bf r},t))
\la{eqn:2.11}
\ee
where ${\hat K} \ast \delta \rho({\bf r},t)\equiv 
\int d {\bf r}'\, K({\bf r}-{\bf r}')\delta \rho({\bf r}',t) $ with the kernel   
 $K({\bf r})\equiv \big(\delta({\bf r})/\rho_0+U({\bf r})/T \big)$.
The nonlinear part ${\cal F}(\delta \rho({\bf r},t))$ in (\ref{eqn:2.11}) is the contribution coming from 
the higher (than quadratic) terms of $F_{id}[\rho]$:
\be
 {\cal F}(\delta \rho({\bf r},t)) \equiv \frac{1}{T}\Big( \frac{\delta F_{id}[\rho]}{\delta \rho}
-\frac{\delta F^G_{id}[\rho]}{\delta \rho}   \Big)=
-\sum_{n=2}^{\infty} \frac{1}{n} \big(-\delta \rho ({\bf r},t)/\rho_0 \big)^n 
\la{eqn:2.12}
\ee
where $ F^G_{id}[\rho] = \frac{T}{2\rho_0} \int d{\bf r} \, (\delta \rho ({\bf r},t))^2$ 
is the Gaussian part of $F_{id}[\rho]$.
If one entirely neglects ${\cal F}(\delta \rho({\bf r},t))$ in (\ref{eqn:2.11}), 
which is tantamount to approximating $F_{id}[\rho]$ by its Gaussian part $ F^G_{id}[\rho]$, then
(\ref{eqn:2.11}) becomes linear and consequently the FDR would be preserved  in the RPT  order by order. 
However, $ F^G_{id}[\rho]$ fails to generate solely the linear diffusion since
\be
\nabla \cdot \Big( \rho({\bf r},t) \nabla \frac{\delta F^G_{id}[\rho]}
{\delta \rho({\bf r},t)} \Big) 
=T\nabla^2 \rho({\bf r},t)+
\frac{T}{\rho_0} \nabla \cdot \big( \delta \rho  \nabla \rho \big)
\la{eqn:2.13}
\ee 
The second nonlinear term in (\ref{eqn:2.13}) gives rise to a totally spurious contribution in the RPT, 
which would yield  incorrect nontrivial result even in the absence of  particle interaction 
\cite{mr,abl,sdd}. It is therefore clear  that the full logarithmic form of $F_{id}[\rho]$ is required to obtain 
the correct behavior for the noninteracting case. 

The transformation (\ref{eqn:2.11}) can be properly linearized by introducing new variables into
the action. In particular, we introduce   a new auxiliary  field $\theta({\bf r},t)$ 
\be
\theta ({\bf r},t) \equiv {\cal F} (\delta \rho({\bf r},t)) =
\frac{1}{T} \frac{\del F_{id}[\rho]}{\del \rho ({\bf r},t)}-\frac{\del \rho ({\bf r},t)}{\rho_0}
\la{eqn:2.14}
\ee
Note that (\ref{eqn:2.14}) limits the new variable $\theta({\bf r},t)$ solely to  the
nonlinear part of the transformation. This slight modification of the original ABL approach 
enables us to eliminate all the unphysical features which the ABL theory was stymied by.

Now incorporating the new field (\ref{eqn:2.14}), we obtain the ideal-gas contribution to  the body force as
\bea
\nabla \cdot \Big( \rho \nabla \frac{\delta F_{id}[\rho]}
{\delta \rho} \Big) &=& T\nabla \cdot \Big(\rho \nabla \Big(\frac{\del \rho}
{\rho_0}+\theta \Big)  \Big)  = T\nabla^2 \rho \nonumber \\
&+&  \frac{T}{\rho_0} \nabla \cdot \big( \delta \rho  \nabla \rho \big)
 +\rho_0 T \nabla^2 \theta +T \nabla \cdot \big( \delta \rho  \nabla \theta \big)
\la{eqn:2.15}
\eea
Comparing (\ref{eqn:2.15}) with (\ref{eqn:2.5}), we note that 
the sum of the last three terms in (\ref{eqn:2.15}) should vanish, and 
hence the new field $\theta({\bf r},t)$ plays the role of  {\em eliminating} the surpurious nonlinear term 
$ (T/\rho_0) \nabla \cdot \big( \delta \rho  \nabla \rho \big)$ in (\ref{eqn:2.13}).
This cancellation was vital to obtain the correct dynamic behavior for the noninteracting case 
nonperturbatively \cite{fnote1}. 

The resulting new action ${\cal S}_{new}[[\rho, \hat\rho, \theta, \hat\theta ]]$ can be decomposed 
into its Gaussian part ${\cal S}^G_{new}[[\rho, \hat\rho, \theta, \hat\theta]]$ and non-Gaussian part 
${\cal S}^{NG}_{new}[[\rho, \hat\rho, \theta, \hat\theta]]$:
 \bea 
{\cal S}_{new} [[  \rho, \hat \rho, \theta, \hat\theta ]]
 &\equiv& {\cal S}^G_{new}[[ \rho, \hat \rho, \theta, \hat\theta ]] 
+{\cal S}^{NG}_{new}[[ \rho, \hat \rho, \theta, \hat\theta ]]\nonumber \\
{\cal S}^G_{new}[[ \rho, \hat \rho, \theta, \hat\theta ]] 
&\equiv& \int d{\bf r} \int dt \, \Big\{ i\hat\rho \Big[ \partial_t \rho
-T \nabla^2 \rho -\u{\rho_0 T \nabla^2 \theta}-\rho_0 \nabla^2 {\hat U} \ast \delta \rho \Big]\nonumber \\
 &-& T \rho_0 ( \nabla \hat \rho)^2 +  i\hat \theta \theta \Big\} \nonumber \\
{\cal S}^{NG}_{new} [[ \rho, \hat \rho, \theta, \hat\theta ]] 
&\equiv&\int d{\bf r} \int dt \, \Big\{ i\hat \rho \Big[
 -\nabla \cdot \Big(\delta \rho \nabla {\hat U} \ast \delta \rho \Big)
- \u{\frac{T}{\rho_0} \nabla \cdot \big( \delta \rho \nabla \rho \big)} \nonumber \\
&-& \u{T \nabla \cdot \big( \delta \rho \nabla \theta \big)} \Big] 
- \, T \delta \rho ( \nabla \hat \rho)^2 - i\hat \theta {\cal F}(\delta\rho) \Big\}
\la{eqn:2.16}
\eea
where $\hat\theta({\bf r},t)$ variable appears from the exponentiation of the delta-functional constraint
$\delta \big[ \theta({\bf r},t)-{\cal F}(\rho({\bf r},t))\big]$. 
The new actions $ {\cal S}^G_{new}[[ \rho, \hat \rho, \theta, \hat\theta ]]$ and 
${\cal S}^{NG}_{new}[[  \rho, \hat \rho, \theta, \hat\theta ]]$ are  now {\em separately} invariant under 
the {\em linear} TR transformation 
\bea
 \quad \rho ({\bf r}, -t)& = & \rho ({\bf r}, t), \quad \quad
 \hat \rho ({\bf r}, -t)  =  -\hat \rho ({\bf r}, t) +
 i {\hat K} \ast \delta \rho({\bf r},t)+i\theta({\bf r},t)  \nonumber \\
\theta ({\bf r}, -t) & = & \theta ({\bf r}, t),  \quad \quad
\hat \theta ({\bf r}, -t)  =  \hat \theta ({\bf r}, t)
+i \partial_t \rho ({\bf r},t)
\la{eqn:2.17}
\eea
It is easy to show that  the modulus of the associated transformation
 matrix $O$ is unity ($\det O=-1$). 
Though with the constraint (\ref{eqn:2.14}) the three underlined terms in (\ref{eqn:2.17}) sum to vanish,  
each should be kept in the explicit calculation of RPT since their presence is crucial for 
separate invariance of the actions 
$ {\cal S}^G_{new}[[\rho, \hat \rho, \theta, \hat\theta ]]$ and ${\cal S}^{NG}_{new}[[\rho, \hat \rho, \theta, \hat\theta ]]$ under TR, which  enables one to construct 
the FDR-preserving RPT from these actions. Nontheless, as physically expected,  
the ultimate effect of these three underlined terms should be their cancellation, 
as explicitly demonstrated in the one-loop calculations in our work. 
Note that the linearization (\ref{eqn:2.17}) inevitably generates
the nonpolynomial (logarithmic) nonlinearity $-i\hat\theta {\cal F}(\delta \rho) $ in the new action, while  
the original action (\ref{eqn:2.6}) contains polynomial nonlinearities only. 
That is the price to pay for linearizing the transformation and thereby preserving the FDR in a RPT. 

Now the response function also takes a new form as
\be
R({\bf r},t; {\bf r}',t')=\frac{i}{T} \Big< \delta \rho( {\bf r},t)\,
 {\hat \theta}( {\bf r}',t') \Big>
 \la{eqn:2.18}
\ee
The FDR (\ref{eqn:2.10}) then immediately follows from the linear transformation
(\ref{eqn:2.17}) by taking correlation of the last member of (\ref{eqn:2.17}) 
with $i\delta \rho({\bf r},t)/T$. 

One is now ready to develop a FDR-preserving RPT for the new action (\ref{eqn:2.16}). 
A full account can be found in our recent paper \cite{bkkk0708}. 
The dynamic equations for the correlation and response functions are formally given by
the matrix Schwinger-Dyson equation which involves the self-energies.
These self-energies can be systematically computed from the two-loop and higher-order 
two-particle irreducible diagrams with vertices mapped out of the non-Gaussian part of 
the action (\ref{eqn:2.16}). 
Since in this work we are aiming to describe the dynamics of fluctuations around the 
equilibrium state (we are excluding the crystalline state from the equilbrium state), 
the equilibrium information must enter as input into the theory through 
the initial values of the dynamic correlation functions. 
Therefore it is crucial to provide the correct static input. 
Through this procedure, we are able to write down 
the  dynamic equation for the Fourier transformed density correlation function $C({\bf k},t)$ alone:
\be 
\partial_t C({\bf k},t)
= -\frac{C({\bf k},t)}{\tau_0({\bf k})}
+\int_0^t  ds\, \Big[\Sigma_{\hat\rho \hat\rho}({\bf k},t-s)\frac
{C({\bf k},s)}{S({\bf k})}-\Sigma_{\hat\rho \hat\theta}({\bf k},t-s)\partial_s C({\bf k},s)\Big]
\la{eqn:2.19}
\ee
where $\tau_0({\bf k}) \equiv S({\bf k})/(\rho_0 T k^2)$ is the 'bare' life time, with
$S({\bf k}) $ being the static structure factor, and $\Sigma_{\hat\rho \hat\rho}({\bf k},t)$ and
$\Sigma_{\hat\rho \hat\theta}({\bf k},t) $ are the self-energies. Note that (\ref{eqn:2.19}) is a nonperturbative closed dynamic equation for 
$C({\bf k},t)$ only since the self-energies can be expressed solely in 
terms of  $C({\bf k},t)$. This was demonstrated in the second reference of \cite{bkkk0708} and 
will be exemplified below. 

The one-loop expressions for the two self-energies in (\ref{eqn:2.19}) are given by
\bea
\Sigma_{\hat\rho \hat\rho}({\bf k},t)&=&
\frac{T^2}{2} \int_{\bf q} \Big[V^2({\bf k},{\bf q})-\frac{k^2}{\rho_0}V({\bf k},{\bf q})
\Big] \, C({\bf q},t) C({\bf k}-{\bf q},t), \nonumber \\
\Sigma_{\hat\rho \hat\theta}({\bf k},t)
&=& \frac{T}{2\rho_0^2} \int_{\bf q} V({\bf k},{\bf q}) \, 
C({\bf q},t) C({\bf k}-{\bf q},t), \nonumber \\
V({\bf k},{\bf q})&\equiv& \Big[({\bf k}\cdot {\bf q}) c({\bf q}) 
+({\bf k}\cdot ({\bf k}-{\bf q})) c({\bf k}-{\bf q})\Big]
\la{eqn:2.20}
\eea
where $c({\bf k})=1/\rho_0-1/S({\bf k})$ is the direct correlation function of the system.
A salient feature of the present theory is that due to 
the correct static input, the bare interaction potential $U({\bf r})$ entirely cancels out, and 
the direct correlation function naturally emerges in the one-loop expression (\ref{eqn:2.20}).

Another intriguing aspect of the this theory is the fact that the nonperturbative equation 
(\ref{eqn:2.19}) takes a distinct form from that of the standard MCT in its structure: the first term in
the convolution integral in (\ref{eqn:2.19}) does not involve the time derivative of 
$C({\bf k},t)$. Moreover, it is very likely due to this fact that the equation (\ref{eqn:2.19}) 
with the one-loop self-energies (\ref{eqn:2.20}) becomes unstable in the long time region.
This kind of situation has been encountered  in the  mode coupling  approach to the dissipative
stochastic systems. This undesirable feature can be eliminated by noting that the conventional 
memory function can be reexpressed in terms of the so-called irreducible memory function \cite{cichohess,kk9597}.
 In particular, the projection operator approach leads to the exact dynamic equation for the correlation 
function of the chosen slow variable $A(t)$ as
\bea
\partial_t C_A(t)&=& -|E_A| C_A(t)+\int_0^t ds \,  M_A(t-s) C_A(s), \nonumber \\
C^L_A(z) &=& C_A(0) \Big[z+|E_A|-M^L_A(z) \Big]^{-1}
\la{eqn:2.21}
\eea
where $|E_A|$ represents a  short time relaxation rate in the system, and $M_A(t)$ is the 
conventional memory function. In (\ref{eqn:2.21}), $C^L_A(z)$ etc  are the Laplace transforms of 
$C_A(t)$ etc: $C^L_A(z)=\int_0^{\infty} dt \, e^{-z t} C_A(t)$, etc. 
The memory function $M_A(t)$ turns out to be further reducible to the   
irreducible memory function $M^{irr}_A(t)$: 
\bea
M_A(t) &=&  M^{irr}_A(t)-|E_A|^{-1} \int_0^t ds\, M_A(t-s) M^{irr}_A(s), \nonumber \\
M^L_A(z) &=& \frac{M^{L,irr}_A(z)}{1+|E_A|^{-1}M^{L, irr}_A(z)}
\la{eqn:2.22}
\eea
 Note that $M_A^{L,irr}(z=0)$ can grow indefinitely when the global relaxation time grows indefinitely 
in contrast to $M_A^L(z=0)$. The above two eqs. lead to the dynamic eq. for $C_A(t)$ 
\bea
\partial_t C_A(t) &=& -|E_A| C_A(t)-|E_A|^{-1} \int_0^t ds\,  M^{irr}_A(t-s) \dot C_A(s), \nonumber \\
C^L_A(z) &=& C_A(0) \Big[z+\frac{|E_A|}{1+|E_A|^{-1} M^{L,irr}_A(z)} \Big]^{-1}
\la{eqn:2.23}
\eea
For dissipative systems with detailed balance, while the mode coupling approximation directly 
applied to the ususal memory kernel $M_A(t)$ in 
 (\ref{eqn:2.21}) often leads to unphysical results, the same approximation applied 
for the irreducible memory kernel $M_A^{irr}(t)$ in (\ref{eqn:2.23}) always yieds physically sensible 
results. See \cite{kk9597}. 

Adopting the irreducible memory function approach, we rewrite (\ref{eqn:2.19}) into the form
\be
\partial_t C({\bf k},t)= -\frac{\rho_0 T k^2}{S({\bf k})} C({\bf k},t)
-\int_0^t  ds\, {\cal M}({\bf k},t-s)\partial_s C({\bf k},s)
\la{eqn:2.24}
\ee
The irreducible memory kernel ${\cal M}({\bf k},t)$ then obeys the following equation \cite{fnote2}: 
\be
{\cal M}({\bf k},t) = \tilde \Sigma_{\hat\rho \hat\rho}({\bf k},t) +\Sigma_{\hat\rho \hat\theta}({\bf k},t)
+\int_0^{t} ds \,  {\cal M}({\bf k},t-s)\tilde \Sigma_{\hat\rho \hat\rho}({\bf k},s)
\la{eqn:2.25}
\ee
where $ \tilde \Sigma_{\hat\rho \hat\rho}({\bf k},t)  \equiv \Sigma_{\hat\rho \hat\rho}({\bf k},t)/(\rho_0Tk^2)$.
Note then that the sum of the two self-energies, that is, the first and second terms on RHS of (\ref{eqn:2.25})
yields the standard mode coupling kernel: 
\bea
&&\Sigma_{MC}({\bf k},t) \equiv \tilde \Sigma_{\hat\rho \hat\rho}({\bf k},t) 
+ \Sigma_{\hat\rho \hat\theta}({\bf k},t) \nonumber \\
 &=&  \frac{T}{2\rho_0} \int_{\bf q} \Big[(\hat{\bf k}\cdot {\bf q}) c({\bf q}) 
+(\hat{\bf k}\cdot ({\bf k}-{\bf q})) c({\bf k}-{\bf q})\Big]^2  C({\bf q},t) C({\bf k}-{\bf q},t)
\la{eqn:2.26}
\eea
where $\hat {\bf k} \equiv {\bf k}/k$. 
Now, when ${\cal M}({\bf k},t)$ is obtained by solving (\ref{eqn:2.25}) iteratively, 
the convolution integral generates the terms $\int_0^{t} ds \, \Sigma_{MC}({\bf k},t-s)\tilde
\Sigma_{\hat\rho \hat\rho}({\bf k},s)+\cdots$. All these terms belong to the higher-loop orders and can be 
dropped in the one loop order, that is, ${\cal M}({\bf k},t)\rightarrow \Sigma_{MC}({\bf k},t)$ 
retains the one-loop two-particle irreducible structure. 
It is therefore perfectly legitimate up to the one-loop order to retain the first two terms, 
ignoring the terms generated by the convolution integral on RHS of (\ref{eqn:2.25}). 
Substituting ${\cal M}({\bf k},t)=\Sigma_{MC}({\bf k},t)$ into (\ref{eqn:2.24}), one recovers 
the standard MCT equation: 
\be
\partial_t C({\bf k},t)
= -\frac{\rho_0 T k^2}{S({\bf k})} C({\bf k},t)
-\int_0^t  ds\, \Sigma_{MC} ({\bf k},t-s)\partial_s C({\bf k},s) 
\la{eqn:2.27}
\ee
We emphasize again that the bare potential completely cancel out in (\ref{eqn:2.27})
 with (\ref{eqn:2.26}) as well as in the previous equation (\ref{eqn:2.19}) due to the correct static 
input relation. We believe that this cancellation is no accident and can be traced back to our 
use of the Legendre-transformed vertex functional where bare interaction potential no longer appears explicitly. 

Here we discussed how to construct a FDR-consistent field theory for a dynamic density 
functional model applicable to colloids as well as to atomic liquids, with emphasis on
the underlying structure of the theory. The time-reversal symmetry of the action 
integral played a vital role in developing such a field theory. 
The consistency with FDR is guaranteed when the time-reversal field transformation
is linearized through the introduction of a new set of auxiliary field variables.
The renormalized perturbation theory is then developed for the new action incorporating
these new variables. Reappearance of the logarithmic nonlinearity in the new action is the price 
one has to pay for the linearization. 

For noninteracting particle systems we recover a simple diffusion law as one expects, which 
is not the case in some recent works \cite{mr,abl,sdd}. 
For interacting particle cases, the dynamic equation for the density correlation function 
with one-loop self-energies is not only different from that of the standard MCT, 
but also is likely to be subject to the long-time instability.  
The irreducible memory function approach enables us to  recover the standard MCT 
in the one-loop order of the renormalized perturbation theory. 
Another salient feature of our theory is the full cancellation of the bare interaction 
potential and the emergence of the direct correlation function which embodies the effect of
the interaction potential, due to the correct static input.

The present theory can be extended to several directions. We can first attempt to 
perform higher-order loop calculations. For this case, however, since the present renormalized perturbation
theory is not an expansion in terms of the smallness parameter, 
it is much desirable to find such a smallness parameter in order for the higher-loop calculations 
to be really meaningful. Despite some difficulties already encountered before \cite{kk03}, there is a possibility
one can develop such a perturbation theory for system with the Kac-type long-range interaction \cite{kkloh}
employing the inverse force range as a smallness parameter. 
It would be also interesting to extend the theory to the inhomogeneous system \cite{yamaguchi05,bbmr}
under an external field where the spatial translation invariance breaks down. 
The calculation of the multibody (like $4$-point) correlation functions \cite{bb07,bbbkmr}
can also be carried out within the present field theory formulation.
We can also attempt to extend our approach to the genuine nonequilibrium problems such as 
ageing dynamics \cite{latz} and to dynamics of colloids under external shear flow \cite{fuchs,mry}. 

One profound problem related to the field theory approach to the glassy dynamics is  
 how to incorporate the thermal activation processes into the present MSR-type field theory.
 Actually the dynamic density functional equation (\ref{eqn:2.1}) 
 was proposed \cite{kk94} as a new mesoscopic kinetic equation which 
 contains not only the nonlinear feed back mechanism leading to the standard MCT, but also can
 naturally incorporate the thermal activation processes.
 The latter statement is based on the argument that the only stationary solution of 
the FP equation (\ref{eqn:2.4}) is the equilibrium state, provided that the FP operator is nonsingular 
\cite{kkbk0102}. 
That is, the system will certainly remain ergodic since it will relax to the equilibrium state.
Yet our one-loop analysis totally misses thermal activation, which indicates that
thermal activation is a nonperturbative process. 
So far we have only used the uniform stationary state of the action in our loop expansion.
We may have to consider some symmetry-breaking stationary configuration of the action
around which a loop expansion can be made. It is a truly challanging problem to incorporate thermal activation 
process within a MSR-type field theory.

\section*{Acknowledgements}
We gratefully acknowldege invaluable discussions with Prof. Kunimasa Miyazaki. 
KK was supported by Grant-in-Aid for Scientific Research (C) Grant 17540366 by Japan Society for the Promotion 
of Science (JSPS). BK was financially supported during his sabbatical stay by the Institute for Molecular Science.

%


\begin{thebibliography}{99}
\bibitem{fnote1}  A statement is made in \cite{br07} that our demonstration of the diffusion law 
for non-interacting systems ignores vertex corrections. 
However our demonstration  in \cite{bkkk0708} is a nonperturbative  result 
independent of loop expansion.
\end{thebibliography}
\end{document}